# Electronic vs. phononic thermal transport in Cr-doped V$_2$O$_3$ thin films across the Mott transition

Johannes Mohr,[1,a)] Kiumars Aryana,[4] Md. Rafiqul Islam,[4] Dirk J. Wouters,[1] Rainer Waser,[1,2] Patrick E. Hopkins,[4,5,6] , Joyeeta Nag,[3] Daniel Bedau[3,a)]

[1] Institute of Materials in Electrical Engineering and Information Technology II, RWTH Aachen University, 52074 Aachen, Germany
[2] Peter Grünberg Institute PGI 7, Research Center Jülich, 52428 Jülich, Germany
[3] Western Digital San Jose Research Center, 5601 Great Oaks Parkway, San Jose, CA 95119, USA
[4] Department of Mechanical and Aerospace Engineering, University of Virginia, Charlottesville, VA 22904, USA
[5] Department of Materials Science and Engineering, University of Virginia, Charlottesville, VA 22904, USA
[6] Department of Physics, University of Virginia, Charlottesville, VA 22904, USA
[a)] Authors to whom correspondence should be addressed: mohr@iwe.rwth-aachen.de, daniel.bedau@wdc.com

**Abstract**

Understanding the thermal conductivity of chromium doped V$_2$O$_3$ is crucial for optimizing the design of selectors for memory and neuromorphic devices. We utilized the time-domain thermoreflectance technique to measure the thermal conductivity of chromium doped V$_2$O$_3$ across varying concentrations, spanning the doping induced metal-insulator transition. In addition, different oxygen stoichiometries and film thicknesses were investigated in their crystalline and amorphous phases. Chromium doping concentration (0%-30%) and the degree of crystallinity emerged as the predominant factors influencing the thermal properties, while the effect of oxygen flow (600-1400 ppm) during deposition proved to be negligible. Our observations indicate that even in the metallic phase of V$_2$O$_3$, the lattice contribution is the dominant factor in thermal transport with no observable impact from the electrons on heat transport. Finally, the thermal conductivity of both amorphous and crystalline V$_2$O$_3$ was measured at cryogenic temperatures (80-450 K). Our thermal conductivity measurements as a function of temperature reveal that both phases exhibit behavior similar to amorphous materials, indicating pronounced phonon scattering effects in the crystalline phase of V$_2$O$_3$.

Devices based on chromium-doped V$_2$O$_3$[1,2] (Cr:V$_2$O$_3$), a prototypical Mott-insulator, have recently received considerable attention. These devices are promising for applications such as selector devices in dense RRAM crossbar memories[3] and in neuromorphic[4] or reservoir computing.[5] Additionally, they are of fundamental physical interest due to the potentially electronically triggered Mott-transition[6–8] they might exhibit. A proposed explanation for the switching effect in these materials is a thermal runaway caused by an applied voltage,[4,9] much like the one previously observed in NbO$_2$.[10] However, alternative explanations more directly related to the Mott-insulating state have also been proposed.[6,7] A key roadblock limiting the advancement of these devices is the lack of materials knowledge that is required for a rational design strategy. While the electrical properties of Cr:V$_2$O$_3$ are reasonably well understood, the thermal properties remain less explored. No studies have investigated the thermal conductivity of chromium doped $V_2O_3$, with only one study focusing on single crystals of undoped $V_2O_3$.[11] This gap in knowledge is particularly concerning, because in the case of the metal-insulator transition (MIT) in the related oxide $VO_2$, significant deviations from the expected behavior have been reported,[12] with the most striking being a violation of the Wiedemann-Franz law in the metallic phase.[13] In response, we conducted an investigation into the way the thermal conductivity of $Cr:V_2O_3$ thin-films evolves over the doping driven Mott-transition, studying apart from the doping concentration, also the influence of stoichiometry and film thickness.

The phase diagram of crystalline $Cr:V_2O_3$, shown in **Figure 1** (a) consists of three distinct domains: the paramagnetic metal (PM) and paramagnetic insulator (PI) regions at room temperature, and an antiferromagnetic insulating (AFI) phase at low temperatures. The transition from the PM to the PI region is caused by chromium doping concentration larger than approximately 1.5% and can be reversed by the application of hydrostatic pressure.[1] Approximately 4 kbar are required per percent chromium to trigger this reversal. This transition is particularly significant as it represents a prototypical Mott-transition,[1] and addresses the impracticality of operating electronic devices at cryogenic temperatures. The transition temperature to the AFI phase is about 155 K for undoped $V_2O_3$,[14] and can be controlled by both Cr and Ti doping.[15] In addition to crystalline films, we have also included amorphous $Cr:V_2O_3$ in our study. This is because there has been considerable interest[16,17] in this material for device applications, and because very little is known about its properties in general. In contrast to the crystalline form, no phase transitions are expected in the amorphous case.

The outline of this work is as follows: First, we describe the samples, their fabrication and the measurement procedure. Then we report on the thermal properties in three steps. To clarify how thin films differ from the bulk material we investigate the behavior of undoped $V_2O_3$. This also allows us to show that the influence of the oxygen stoichiometry is negligible. Next, we characterize the influence of different levels of chromium doping in the crystalline and amorphous case, which corresponds to a crossing of the PM-PI phase boundary. Finally, again for undoped $V_2O_3$, we report on the behavior at different cryogenic temperatures, which yields data for the PM-AFI transition.

The samples investigated were thin films with thicknesses ranging between 10 nm to 120 nm. They were fabricated on silicon wafers covered with a 5 nm thick TiN layer, which were cleaned by an ultrasonic treatment in acetone and isopropanol for 10 min each. After sonication, the samples were thoroughly rinsed in deionized water and dried with nitrogen. The $Cr:V_2O_3$ films were then deposited by reactive radio-frequency magnetron sputtering from 1" targets that consisted of an alloy corresponding to the desired composition. The process gas was an argon/oxygen mixture containing either 600 ppm or 1400 ppm $O_2$; this leads to the formation of the correct $V_2O_3$ phase. Depending on whether an amorphous or crystalline film was desired, the depositions were done at room temperature or at 600°C, respectively. The deposition process has been extensively characterized in previous works,[18] verifying the correct composition and structure of the films. Furthermore, we recently demonstrated that the pressure or doping driven MIT is maintained even for very thin films.[19] To prevent oxidation and contamination of the films, the samples were then capped in-situ with a 10 nm thick Pt layer. Finally, an 80 nm Ru transducer layer was deposited. The resulting $TiN/Cr:V_2O_3/Pt$ stack was selected because this is identical to the one used for nanoscale selector devices,[3,4] and therefore, the measured thermal conductivities should carry over well to the simulation of devices. Doping concentrations between 0% and 30% were investigated. The location of these compositions in the phase diagram is shown in Figure 1(a).

Finally, undoped $V_2O_3$ films were also fabricated with 10 nm Pt bottom electrodes for use in temperature dependent measurements, resulting in a Pt/$V_2O_3$/Pt stack. This platinum layer is identical to the one on top, grown directly on a silicon substrate. For the crystalline process these exhibit a rather rough bottom interface due to some formation of PtSi, but still data could be acquired. The amorphous process should not be affected by this issue.

The thermal conductivity of all the thin films studied here was measured using time-domain thermoreflectance (TDTR), a well-established technique extensively discussed in prior works.[20–23] Briefly, relevant to the results presented in this paper, the output of an 80 MHz Ti laser is split into pump and probe paths. The pump path is modulated at a frequency of 8.4 MHz, while the probe path is delayed in time relative to the pump beam using a mechanical delay stage. The pump and probe beams are redirected, overlapped using a dichroic mirror, and focused to spot sizes of 20 μm and 12 μm, respectively. Changes in the reflectivity of the sample are captured using a balanced photodiode paired with a lock-in amplifier. The surface of the samples is coated with an 80 nm specular metallic layer of ruthenium (Ru) to serve as a transducer, facilitating the detection of thermoreflectance variations on the sample surface due to thermal excursions in the underlying materials. Given the low thermal conductivity of $V_2O_3$ and the large beam diameter relative to the film thickness, the measurements are predominantly sensitive to cross-plane thermal conductivity.

In the sample configuration used in this study (Ru/Pt/$V_2O_3$/TiN/Si), the presence of multiple interfaces and layers with unknown thermal conductivities makes it challenging to isolate the intrinsic thermal conductivity of Cr:$V_2O_3$. This difficulty arises because single TDTR measurements only provide the overall thermal resistance between the Ru transducer and the Si substrate. By knowing the interfacial thermal conductance and thermal conductivity of other layers, one can theoretically extract the thermal conductivity of an unknown layer. However, with several unknown layers, an alternative approach is required to measure the intrinsic thermal conductivity of the layer of interest. To enable this, a series of otherwise identical stacks with different Cr:$V_2O_3$ thicknesses is fabricated. The total (area referred) thermal resistance $R_{th}$ can be written as

$$R_{th}(l) = \frac{1}{k}l + R_{fix} \qquad (1)$$

with the film thickness $l$ and the thermal conductivity $k$ of Cr:$V_2O_3$, as well as the fixed contribution of all interfaces and electrodes $R_{fix}$. Because of this, $k$ can be obtained from a simple linear regression.[24,25] This is only possible if the film properties, and thus $k$, are independent of the layer thickness, an assumption which must be verified by inspecting whether the relation between $k$ and $l$ is in fact linear.

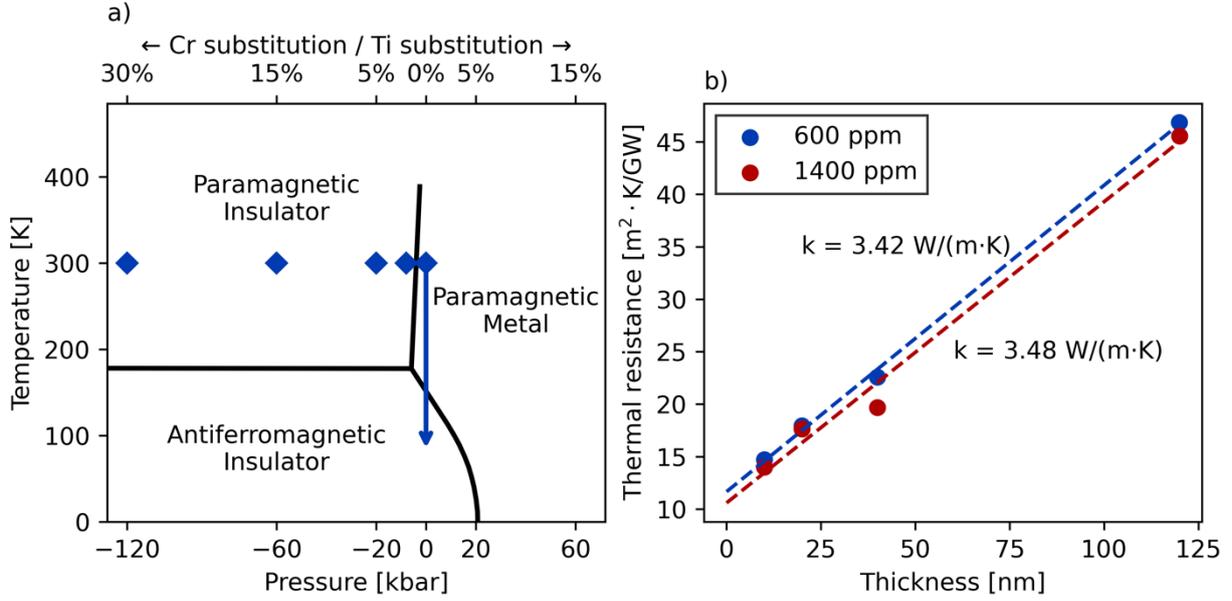

**Fig. 1.** (a) The schematic phase diagram of doped $V_2O_3$. Negative pressures indicate that the material must be compressed to return it to a state equivalent to undoped $V_2O_3$. Diamonds show the characterized compositions; the arrow denotes the temperature dependent measurement for undoped films. (b) Measured thermal resistances for different film thicknesses of undoped $V_2O_3$. The difference between a low (600 ppm) and high (1400 ppm) oxygen content in the films is negligible.

We begin the discussion of our results with the behavior of undoped polycrystalline $V_2O_3$ in thin film form, before turning to the influence of chromium doping. This is because the properties of thin films can deviate significantly from bulk crystals.[26] Figure 1(b) shows the measured thermal resistance over film thickness for $V_2O_3$ films fabricated under two different oxygen flows. As one would expect, the thermal resistance of the stack between Ru and Si should linearly increase with increasing the thickness of $V_2O_3$. Therefore, values for $k$ can be extracted from Eq. 1, yielding 3.42 W/(m·K) for 600 ppm $O_2$ and 3.48 W/(m·K) for 1400 ppm. This is in good agreement with the results of Andreev et al., who found a value of approximately 4.8 W/(m·K) for bulk single crystals, which are expected to have a higher thermal conductivity than polycrystalline films.[11] While the thermal conductivity is reduced in $V_2O_3$ thin films, as is typical,[27,28] about 70% of the bulk value is retained. This could be an indication of a high-quality film growth. The difference between the oxygen concentrations appears to be negligible, which is interesting, because the electrical conductivity is very sensitive to even minor changes in the stoichiometry.[18,29] This would enable tuning the ratio between electrical and thermal conductivity according to the requirements of the application. Furthermore, this finding is a good indication that even in the metallic phase, the electronic contribution to the thermal conductivity $k_e$ is quite small. If a change in stoichiometry leads to a large change in electrical conductivity, this will be accompanied by a corresponding large change in $k_e$. Thus, the total thermal conductivity should change significantly, too, which is not observed. This is only possible if $k_e$ is small in comparison to $k$, which is consistent with the data for single crystals, where an electronic contribution of less than 25% was found.[11]

Interestingly it appears like both 40 nm films lie slightly below the trend. Most likely, this thickness is not truly different, but it is an effect of the order in which the films were deposited. The depositions were done in a randomized order, and by chance, these two were the first runs of a day. This indicates that there could be a slight warm-up effect of the system.

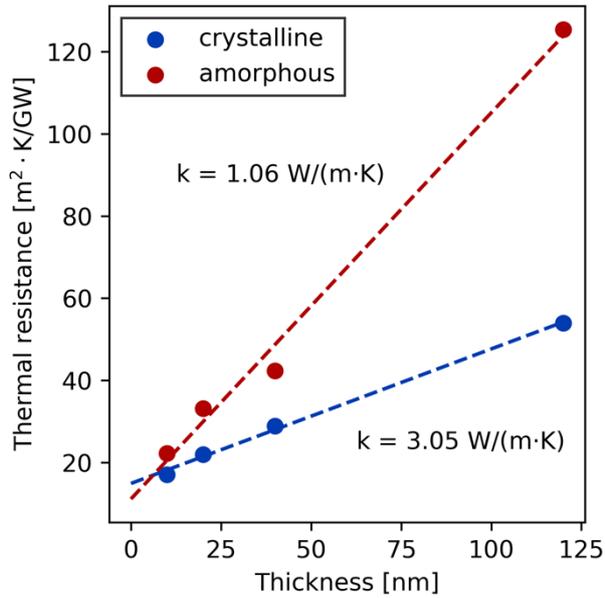

**Fig. 2.** Thermal resistances of 15% doped Cr:$V_2O_3$ films in the crystalline and amorphous phases. Lines indicate the fits used to extract the thermal conductivity.

Having characterized the undoped case, we turn to chromium-doped $V_2O_3$: The results from a thickness series for Cr-doped $V_2O_3$ are shown in **Figure 2**. A concentration of 15% Cr was selected for these measurements, because this composition is the one most studied for devices.[3,4] For the same reason, for all doped films a gas flow of 600 ppm $O_2$ was used. Again, in both the crystalline (Figure 2(a)) and amorphous (Figure 2(b)) case, the assumption of a thickness independent thermal conductivity seems to be approximately satisfied. A value of 3.05 W/(m·K) for the crystalline material compared to 1.06 W/(m·K) for the amorphous one is obtained. Interestingly, in the crystalline case, only a reduction of approximately 0.4 W/(m·K) is found when comparing the material to the undoped case. As this change in composition represents a transition from the PM region deep into the PI region, a much larger difference might be expected. Most likely, this is another indication that the electronic contribution $k_e$ even in the PM phase is small. The smaller value in the amorphous case is plausible due to the stronger disorder, and thus probably smaller lattice contribution.

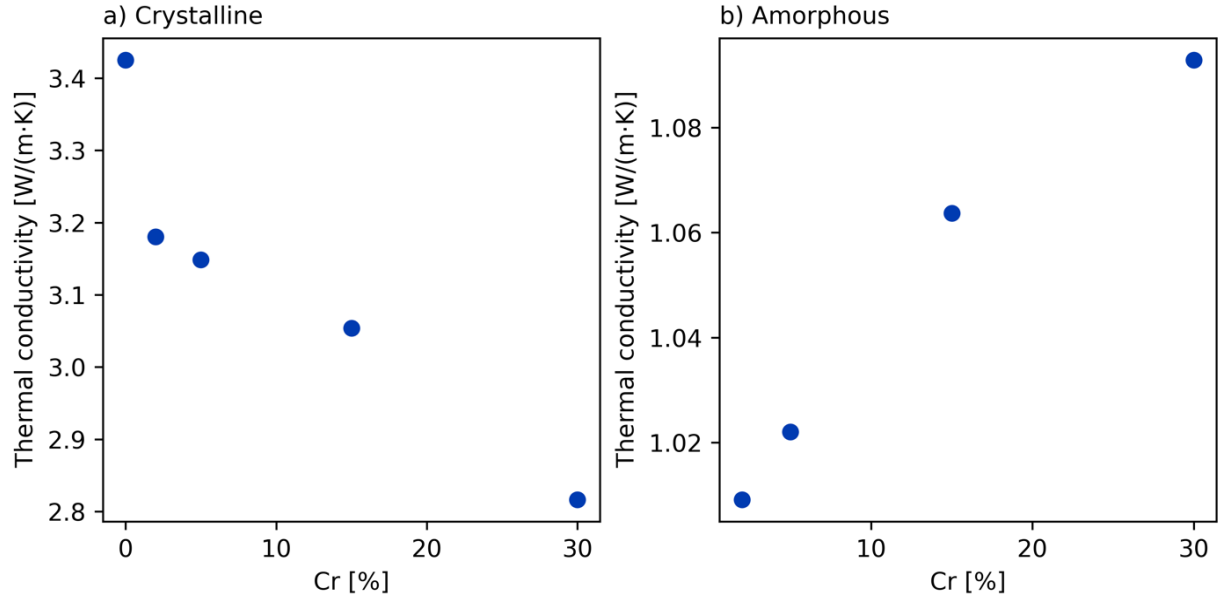

**Fig. 3.** Thermal conductivities for different doping concentrations in the crystalline (a) and amorphous (b) case.

In **Figure 3**, the results for different doping concentrations are reported. These values were obtained from measurements on 120 nm films with the assumption that the influence of interfaces and the electrodes is the same for all doped films, and equal to that measured for 15% doping. This was necessary because it was unfortunately not feasible to deposit and characterize a full thickness series for each concentration. For the undoped material the thermal conductivity obtained from Figure 1 is reported. In both the crystalline and amorphous case, a gradual dependence on the doping concentration is found, except for a small gap of about 0.2 W/(m·K) from the undoped to the 2% doped crystalline samples.

Interestingly, opposite trends are observed here, with the thermal conductivity decreasing for crystalline films, and slightly increasing for amorphous ones. In any case, this dependence is relatively weak, with only a change of about 10% over the measured range, which is close to the measurement limit. In the crystalline case, this behavior appears plausible. The change in the lattice contribution can be assumed to be from to an increase in impurity scattering due to the dopants. The scattering rate on substitutional dopants typically depends mainly on the mass difference,[30] and because the atomic masses of vanadium and chromium are approximately 51 and 52, respectively, these rates will be very small. In this case, the addition of more chromium would only induce a small decrease in thermal conductivity.

An estimate of $k_e$ can be obtained from the electrical conductivity and the Wiedemann-Franz (WF) law. This is shown in **Figure 4** with literature values for both single crystals and thin films. Clearly, even for thin films, which are low resistive compared to single crystals, any contribution will be negligible for doping concentrations over 5%. Even for 2% and below, $k_e$ will be only a small fraction of the total thermal conductivity.

On the other hand, if the 2% doped films are regarded as nearly electrically insulating and the undoped ones as conductive, Figure 3 implies an additional $k_e$ of approximately 0.2 W/(m·K) for undoped films. That would mean a quite high electrical resistivity of $3.6 \times 10^{-2}$ Ω·cm compared to the references. This is consistent with some earlier measurements on our films,[18] however their electrical properties proved to be strongly thickness-dependent, which makes a direct comparison difficult. An additional issue is that electrical measurements of the undoped low resistance films are typically done in-plane in a van-der-Pauw configuration, while the thermal measurements were done through-plane. Therefore, it seems questionable to give a definitive value for $k_e$, but it can be reliably concluded that it is only a small part of the total $k$.

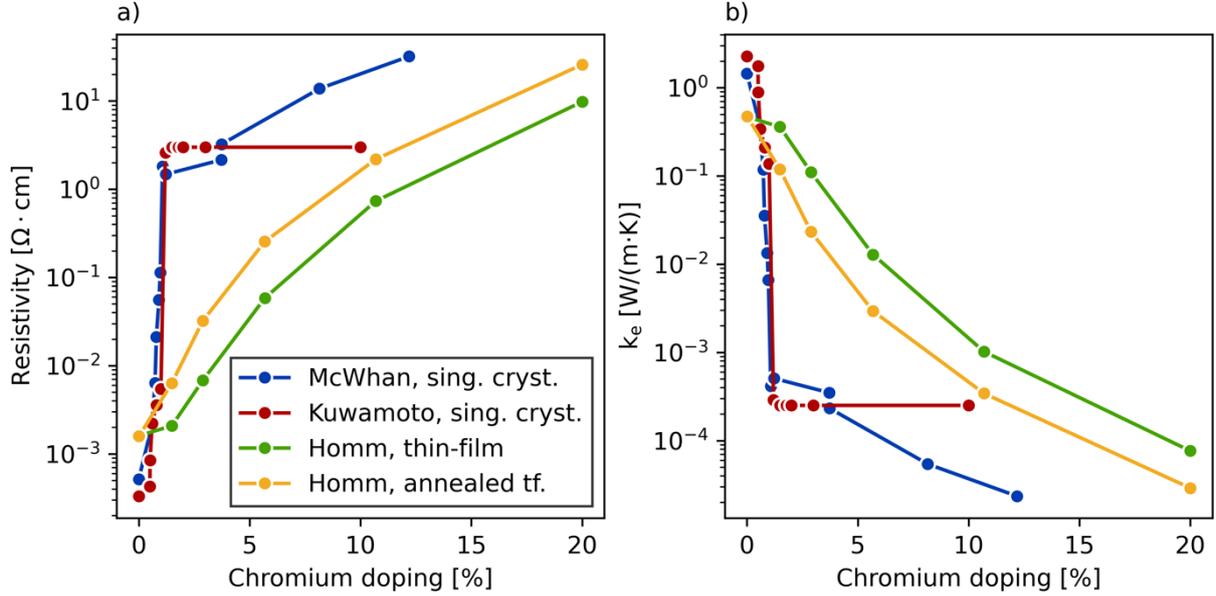

**Fig. 4.** (a) Literature values for the electrical resistivities of Cr:$V_2O_3$ for both single crystals[1,31] as well as thin-films.[32] The latter are shown both before and after an anneal. (b) The electronic contribution to the thermal conductivity, assuming $L = 2.44 \times 10^{-8}$ $V^2K^{-2}$.

The fact that $k$ varies relatively little with doping, while the electrical conductivity changes over orders of magnitude, is quite important for device applications. It makes it possible to optimize both properties separately. For example, an additional low doped layer could be added to the device stack at the interfaces to the electrodes. Such a layer would thermally insulate the electrodes from the highly doped switching layer, thereby minimizing heat losses out of the cell. Electrically, the behavior would remain largely unchanged. Fabrication of the active and insulating layer from the same material will greatly simplify process integration because issues such as interdiffusion, adhesion, etc. can be avoided.

Finally, we have also performed an investigation into the effect of temperature in undoped $V_2O_3$. This is both to characterize the behavior of $k$ across PM-AFM phase transition, as well as to get a general idea about the direction and magnitude of its temperature dependence above the transition temperature. Measurements were performed on 20 nm films to avoid the need to take into account the heat capacity, for which values are not readily available at low temperatures. Furthermore, the effective thermal conductivity

$$k_{\text{eff}} = \frac{l}{R_{\text{th}}} \qquad (2)$$

is reported which includes both interfaces and the films themselves. This is to avoid making strong assumptions about the change of the interfacial and electrode thermal resistance with temperature. The results are shown in **Figure 5** (a).

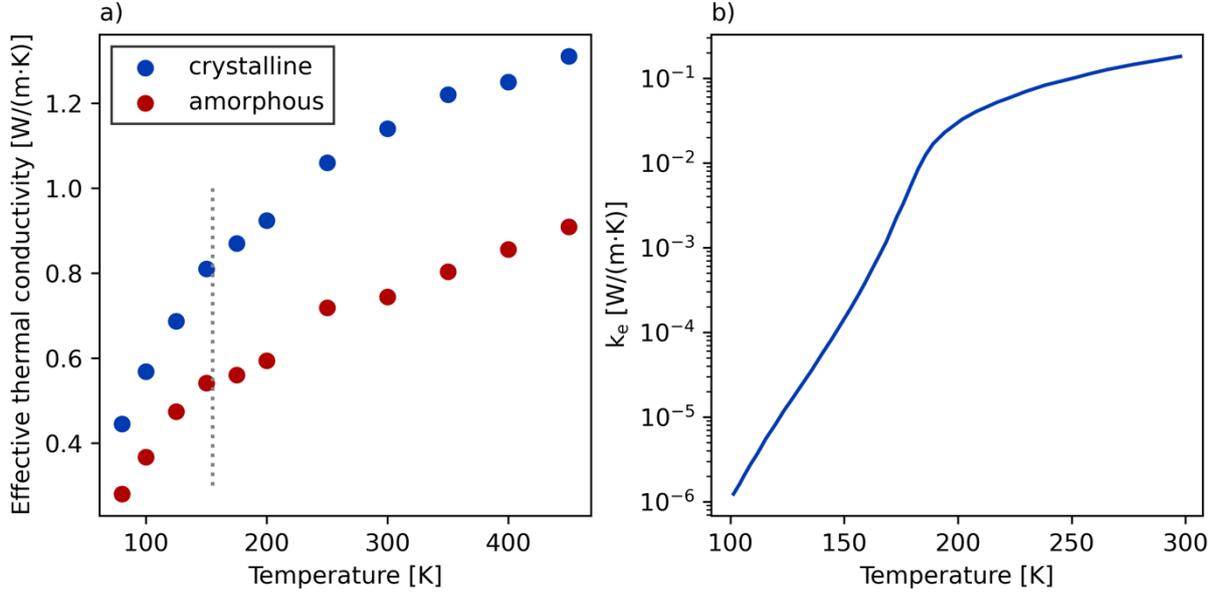

**Fig. 5.** (a) Effective thermal conductivities of 20 nm thin films of undoped crystalline and amorphous $V_2O_3$ films. (b) The electronic contribution to the thermal conductivity derived from electrical resistivities for a comparable film (as measured by Rupp[18]).

Clearly, $k$ increases with temperature and there appear to be two regions with different slopes. The change in slope seems to happen at approximately 155 K, which corresponds to the temperature-driven phase transition.[14] This is somewhat contrary to the earlier findings of Andreev et al. on single crystals. They too report an increase above the transition temperature,[11] but they find the opposite behavior below, with $k$ decreasing with temperature. Our results might at first glance be attributed to the imperfect interfaces of the film grown on platinum, but these should only affect the crystalline film. The amorphous samples however show the same behavior. This is quite surprising, because no phase transition would be expected in this case, yet the change in slope appears to be even more pronounced. It might be speculated that the material is in fact nano-crystalline and therefore shows a similar behavior, but the present data do not allow for a definitive conclusion.

According to Figure 5(a), only a moderate change in the thermal conductivity is observed across the low temperature MIT. Given that the electrical resistivity changes by several orders of magnitude, a large change in $k_e$ could be expected. However, based on our observations for thin film $V_2O_3$, the resistivity of the metallic phase at room temperature is still relatively high, which leads to a small contribution from electrons. Figure 5(b) shows $k_e$ calculated from four-point measurement data[18] acquired on a film comparable to the crystalline one in Figure 5(a). The maximum estimated contribution from electrons in thermal conductivity is about 0.18 W/(m·K) and thus only about 5% of the value determined for the undoped films in Figure 4(b), which is within uncertainty of TDTR measurements. Based on our electrical resistivity measurement and WF calculations, we can conclude that in Cr-doped $V_2O_3$ only vibrational modes from the lattice are contributing to the thermal conductivity in both insulating and metallic phases. This is also observed in chalcogenide-based phase change materials, where orders of magnitude changes in electrical resistivity only alter the electron contribution for specific compositions.[33]

To summarize, we have characterized the thermal conductivity of thin chromium doped $V_2O_3$ films as a function of parameters such as the doping concentration, oxygen stoichiometry and film thickness, which strongly influence the electrical resistivity. Both the stoichiometry and the film thickness have a negligible influence on the thermal conductivity. The doping concentration only changes by at most 10%. In the PM phase, an electronic contribution to $k$ can be observed, but like in single crystals,[11] this is only a small part of its total value. The lattice thermal conductivity is dominating in all regions of the phase diagram. Finally, we found that the behavior of *k* over the transition to the AFI phase appears significantly different from the bulk case; and discovered a so far not explained similar transition behavior for amorphous films.

**Acknowledgements**

We appreciate support from the National Science Foundation, Grant No. 2318576 and Air Force Office of Scientific Research, Grant No. FA9550-22-1-0456.
JM, DJW and RW acknowledge funding by the DFG (German Science Foundation) within the collaborative research center SFB 917.

**Conflict of Interest Statement**

The authors have no conflicts to disclose.

**Author Contributions**

**Johannes Mohr:** Writing – original draft (lead); Investigation (equal); Data curation (equal); Formal analysis (lead); Visualization (lead). **Kiumars Aryana:** Writing – review and editing (equal); Investigation (equal); Data curation (equal). **Md. Rafiqul Islam:** Writing – review and editing (equal); Investigation (equal); Data curation (equal). **Joyeeta Nag:** Conceptualization (equal); Supervision (equal); Writing – review and editing (equal). **Dirk J. Wouters:** Conceptualization (equal); Writing – review and editing (equal). **Rainer Waser:** Supervision (equal); Funding acquisition (equal); Writing – review and editing (equal). **Patrick E. Hopkins:** Supervision (equal); Writing – review and editing (equal). **Daniel Bedau:** Conceptualization (equal); Supervision (equal); Project administration (equal); Writing – review and editing (equal).

**Data Availability Statement**

The data that support the findings of this study are available from the corresponding author upon reasonable request.

**References**

[1] D.B. McWhan, and J.P. Remeika, "Metal-Insulator Transition in $(V_{1-x}Cr_x)_2O_3$," Phys Rev B **2**(9), 3734–3750 (1970).
[2] D.B. McWhan, T.M. Rice, and J.P. Remeika, "Mott Transition in Cr-Doped V2O3," Phys Rev Lett **23**(24), 1384–1387 (1968).
[3] T. Hennen, D. Bedau, J.A.J. Rupp, C. Funck, S. Menzel, M. Grobis, R. Waser, and D.J. Wouters, "Forming-free Mott-oxide threshold selector nanodevice showing s-type NDR with high endurance (>10\^12 cycles), excellent Vth stability (< 5%), fast (< 10 ns) switching, and promising scaling properties," in *2018 IEEE International Electron Devices Meeting (IEDM), 1-5 December 2018, San Francisco, USA*, (2018 IEEE International Electron Devices Meeting (IEDM), 1-5 December 2018, San Francisco, USA, 2018).

[4] T. Hennen, D. Bedau, J.A.J. Rupp, C. Funck, S. Menzel, M. Grobis, R. Waser, and D.J. Wouters, "Switching Speed Analysis and Controlled Oscillatory Behavior of a Cr-doped V2O3 Threshold Switching Device for Memory Selector and Neuromorphic Computing Application," 2019 Ieee 11th International Memory Workshop (Imw 2019), 44–47 (2019).

[5] W. Ma, T. Hennen, M. Lueker-Boden, R. Galbraith, J. Goode, W.H. Choi, P. Chiu, J.A.J. Rupp, D.J. Wouters, R. Waser, and D. Bedau, "A Mott Insulator-Based Oscillator Circuit for Reservoir Computing," ISCAS2020, 1–5 (2020).

[6] P. Stoliar, L. Cario, E. Janod, B. Corraze, C. Guillot-Deudon, S. Salmon-Bourmand, V. Guiot, J. Tranchant, and M. Rozenberg, "Universal electric-field-driven resistive transition in narrow-gap Mott insulators," Advanced Materials **25**(23), 3222–3226 (2013).

[7] E. Janod, J. Tranchant, B. Corraze, M. Querré, P. Stoliar, M. Rozenberg, T. Cren, D. Roditchev, V.T. Phuoc, M.-P. Besland, and L. Cario, "Resistive Switching in Mott Insulators and Correlated Systems," Adv Funct Mater **25**(40), 6287–6305 (2015).

[8] M. Querré, J. Tranchant, B. Corraze, S. Cordier, V. Bouquet, S. Députier, M. Guilloux-Viry, M.-P. Besland, E. Janod, and L. Cario, "Non-volatile resistive switching in the Mott insulator (V1−xCrx)2O3," Physica B Condens Matter **536**, 327–330 (2018).

[9] T.A. Hennen, Harnessing Stochasticity and Negative Differential Resistance for Unconventional Computation, 2023.

[10] C. Funck, S. Menzel, N. Aslam, H. Zhang, A. Hardtdegen, R. Waser, and S. Hoffmann-Eifert, "Multidimensional Simulation of Threshold Switching in NbO2 Based on an Electric Field Triggered Thermal Runaway Model," Adv Electron Mater **2**(7), (2016).

[11] V.N. Andreev, F.A. Chudnovskii, A. V Petrov, and E.I. Terukov, "Thermal conductivity of VO2, V3O5, and V2O3," Physica Status Solidi (a) **48**(2), K153–K156 (1978).

[12] C.N. Berglund, and H.J. Guggenheim, "Electronic Properties of VO$_2$ near the Semiconductor-Metal Transition," Physical Review **185**(3), 1022–1033 (1969).

[13] S. Lee, K. Hippalgaonkar, F. Yang, J. Hong, C. Ko, J. Suh, K. Liu, K. Wang, J.J. Urban, X. Zhang, C. Dames, S.A. Hartnoll, O. Delaire, and J. Wu, "Anomalously low electronic thermal conductivity in metallic vanadium dioxide," Science (1979) **355**(6323), 371–374 (2017).

[14] F.J. Morin, "Oxides Which Show a Metal-to-Insulator Transition at the Neel Temperature," Phys Rev Lett **3**(1), 34–36 (1959).

[15] D.B. McWhan, J.P. Remeika, T.M. Rice, W.F. Brinkman, J.P. Maita, and A. Menth, "Electronic Specific Heat of Metallic Ti-Doped V2O3," Phys Rev Lett **27**(14), 941–943 (1971).

[16] J.A.J. Rupp, M. Querré, A. Kindsmüller, M.-P. Besland, E. Janod, R. Dittmann, R. Waser, and D.J. Wouters, "Different threshold and bipolar resistive switching mechanisms in reactively sputtered amorphous undoped and Cr-doped vanadium oxide thin films," J Appl Phys **123**(4), 44502 (2018).

[17] J. Mohr, C. Bengel, T. Hennen, D. Bedau, S. Menzel, R. Waser, and D.J. Wouters, "Physical Origin of Threshold Switching in Amorphous Chromium-Doped V$_2$O$_3$," Physica Status Solidi (a), (2023).

[18] J.A.J. Rupp, Synthesis and Resistive Switching Mechanisms of Mott Insulators Based on Undoped and Cr-Doped Vanadium Oxide Thin Films, 2020.

[19] J. Mohr, T. Hennen, D. Bedau, R. Waser, and D.J. Wouters, "Bulk-Like Mott-Transition in Ultrathin Cr-Doped V$_2$O$_3$ Films and the Influence of its Variability on Scaled Devices," Advanced Physics Research, (2024).

[20] D.G. Cahill, "Analysis of heat flow in layered structures for time-domain thermoreflectance," Review of Scientific Instruments **75**(12), 5119–5122 (2004).

[21] K. Kang, Y.K. Koh, C. Chiritescu, X. Zheng, and D.G. Cahill, "Two-tint pump-probe measurements using a femtosecond laser oscillator and sharp-edged optical filters," Review of Scientific Instruments **79**(11), (2008).

[22] P. Jiang, X. Qian, and R. Yang, "Tutorial: Time-domain thermoreflectance (TDTR) for thermal property characterization of bulk and thin film materials," J Appl Phys **124**(16), (2018).

[23] K. Aryana, D.A. Stewart, J.T. Gaskins, J. Nag, J.C. Read, D.H. Olson, M.K. Grobis, and P.E. Hopkins, "Tuning network topology and vibrational mode localization to achieve ultralow thermal conductivity in amorphous chalcogenides," Nat Commun **12**(1), 2817 (2021).

[24] M.S. Bin Hoque, Y.R. Koh, J.L. Braun, A. Mamun, Z. Liu, K. Huynh, M.E. Liao, K. Hussain, Z. Cheng, E.R. Hoglund, D.H. Olson, J.A. Tomko, K. Aryana, R. Galib, J.T. Gaskins, M.M.M. Elahi, Z.C. Leseman, J.M. Howe, T. Luo, S. Graham, M.S. Goorsky, A. Khan, and P.E. Hopkins, "High In-Plane Thermal Conductivity of Aluminum Nitride Thin Films," ACS Nano **15**(6), 9588–9599 (2021).

[25] K. Aryana, J.T. Gaskins, J. Nag, D.A. Stewart, Z. Bai, S. Mukhopadhyay, J.C. Read, D.H. Olson, E.R. Hoglund, J.M. Howe, A. Giri, M.K. Grobis, and P.E. Hopkins, "Interface controlled thermal resistances of ultra-thin chalcogenide-based phase change memory devices," Nat Commun **12**(1), 774 (2021).

[26] J.L. Braun, C.H. Baker, A. Giri, M. Elahi, K. Artyushkova, T.E. Beechem, P.M. Norris, Z.C. Leseman, J.T. Gaskins, and P.E. Hopkins, "Size effects on the thermal conductivity of amorphous silicon thin films," Phys Rev B **93**(14), 140201 (2016).

[27] T. Yamane, Y. Mori, S. Katayama, and M. Todoki, "Measurement of thermal diffusivities of thin metallic films using the ac calorimetric method," J Appl Phys **82**(3), 1153–1156 (1997).

[28] D.G. Cahill, H.E. Fischer, T. Klitsner, E.T. Swartz, and R.O. Pohl, "Thermal conductivity of thin films: Measurements and understanding," Journal of Vacuum Science & Technology A: Vacuum, Surfaces, and Films **7**(3), 1259–1266 (1989).

[29] J.A.J. Rupp, B. Corraze, M.P. Besland, L. Cario, J. Tranchant, D.J. Wouters, R. Waser, and E. Janod, "Control of stoichiometry and morphology in polycrystalline V2O3 thin films using oxygen buffers," J Mater Sci **55**(30), 14717–14727 (2020).

[30] R. Gurunathan, R. Hanus, M. Dylla, A. Katre, and G.J. Snyder, "Analytical Models of Phonon–Point-Defect Scattering," Phys Rev Appl **13**(3), 034011 (2020).

[31] H. Kuwamoto, J.M. Honig, and J. Appel, "Electrical properties of the (V1-xCrx)2O3 system," Phys Rev B **22**(6), 2626–2636 (1980).

[32] P. Homm, L. Dillemans, M. Menghini, B. Van Bilzen, P. Bakalov, C.Y. Su, R. Lieten, M. Houssa, D. Nasr Esfahani, L. Covaci, F.M. Peeters, J.W. Seo, and J.P. Locquet, "Collapse of the low temperature insulating state in Cr-doped V2O3 thin films," Appl Phys Lett **107**(11), (2015).

[33] K. Aryana, Y. Zhang, J.A. Tomko, M.S. Bin Hoque, E.R. Hoglund, D.H. Olson, J. Nag, J.C. Read, C. Ríos, J. Hu, and P.E. Hopkins, "Suppressed electronic contribution in thermal conductivity of Ge2Sb2Se4Te," Nat Commun **12**(1), 7187 (2021).